\documentclass[aps,prc,twocolumn,showpacs,floatfix,nofootinbib,preprintnumbers,superscriptaddress,amsmath,amssymb,longbibliography,amsproc]{revtex4-1}
\usepackage{epsfig,dsfont,amssymb,amsmath,amsthm,amsfonts,amsbsy,mathrsfs}
\usepackage{graphicx}
\usepackage{amsmath}
\usepackage{amssymb}%
\usepackage{dcolumn}
\usepackage{multirow}
\usepackage[colorlinks=true,linkcolor=blue,citecolor=blue,urlcolor=blue]{hyperref}
\usepackage{bbold}
\usepackage{bibentry}

\graphicspath{{.}}

\usepackage{stackengine}

\usepackage[disable]{todonotes}

\newcommand{\be}{\begin{equation}}
\newcommand{\ee}{\end{equation}}
\newcommand{\ba}{\begin{eqnarray}}
\newcommand{\ea}{\end{eqnarray}}

\stackMath
\newcommand\tsup[2][2]{%
 \def\useanchorwidth{T}%
  \ifnum#1>1%
    \stackon[-.5pt]{\tsup[\numexpr#1-1\relax]{#2}}{\scriptscriptstyle\sim}%
  \else%
    \stackon[.5pt]{#2}{\scriptscriptstyle\sim}%
  \fi%
}

\DeclareMathAlphabet{\mathpzc}{OT1}{pzc}{m}{it}

\begin{document}

\title{How to renormalize coupled cluster theory}

\author{Z.~H. Sun}
\affiliation{Physics Division, Oak Ridge National Laboratory, Oak Ridge, Tennessee 37831, USA}

\author{C.~A.~Bell}
\affiliation{Department of Physics and Astronomy, University of Tennessee, Knoxville, Tennessee 37996, USA}

\author{G.~Hagen}
\affiliation{Physics Division, Oak Ridge National Laboratory, Oak Ridge, Tennessee 37831, USA}
\affiliation{Department of Physics and Astronomy, University of Tennessee, Knoxville, Tennessee 37996, USA}

\author{T.~Papenbrock}
\affiliation{Department of Physics and Astronomy, University of Tennessee, Knoxville, Tennessee 37996, USA}
\affiliation{Physics Division, Oak Ridge National Laboratory, Oak Ridge, Tennessee 37831, USA}

\begin{abstract}
Coupled cluster theory is an attractive tool to solve the quantum
many-body problem because its singles and doubles (CCSD) approximation
is computationally affordable and yields about 90\% of the correlation
energy. Capturing the remaining 10\%, e.g. via including triples, is
numerically expensive. Here we assume that short-range three-body
correlations dominate and -- following \citet[“How to renormalize the
  Schr{\"o}dinger equation,” arXiv:nucl-th/9706029]{lepage1997} --
that their effects can be included within CCSD by renormalizing the
three-body contact interaction. We renormalize this contact in
$^{16}$O and obtain accurate CCSD results for $^{24}$O, $^{20-34}$Ne,
$^{40,48}$Ca, $^{78}$Ni, $^{90}$Zr, and $^{100}$Sn.
\end{abstract}


\maketitle

{\it Introduction.---} In the past two decades computations of atomic
nuclei based on Hamiltonians from effective field theories of quantum
chromodynamics have advanced from the lightest nuclei to $^{208}$Pb
~\cite{navratil2004b,forssen2005,nogga2005,hagen2008,epelbaum2011,roth2012,soma2014,contessi2017,bansal2018,arthuis2020,hu2021}.
This progress is based on ideas and insights from effective field
theory~\cite{vankolck1994,epelbaum2009,hammer2020} and the
renormalization group~\cite{bogner2003,bogner2007,tsukiyama2011}, and
on computational solutions of the nuclear many-body problem that are
systematically improvable and scale polynomially with increasing mass
number~\cite{dickhoff2004,lee2009,hagen2010b,elhatisari2015,soma2013,hergert2016,stroberg2017}.

Let us take coupled-cluster
theory~\cite{coester1958,coester1960,cizek1966,kuemmel1978,bishop1991,zeng1998,mihaila2000b,dean2004,bartlett2007,shavittbartlett2009,hagen2010b,binder2013,hagen2014}
as an example. Here, one expresses the ground state as $|\psi\rangle =
e^T|\phi\rangle$, where the reference $|\phi\rangle$ is an $A$-fermion
product state and $T=T_1+T_2+\cdots+T_A$ is a cluster excitation
operator consisting of 1-particle-1-hole (1p-1h) up to $A$p-$A$h
excitations. Its workhorse, the CCSD approximation, truncates
$T=T_1+T_2$ and provides us with an attractive compromise between
accuracy and computational cost. In the Hartree-Fock basis, CCSD
yields about 90\% of the correlation energy (i.e. the difference
between the exact energy and the expectation value
$\langle\phi|H|\phi\rangle$ of the Hamiltonian $H$ in the reference),
while costing an effort that scales as $A^2u^4$ for a single-particle
basis consisting of $A$ occupied and $u$ unoccupied orbitals.

The inclusion of triples excitations, i.e. $T=T_1+T_2+T_3$, typically
yields about 98-99\% of the correlation energy, and similar statements
apply to quantum chemistry~\cite{noga1987,scuseria1988}.  It is not
well understood why triples account for about 10\% of the CCSD
correlation energy~\cite{kutzelnigg1991}, but size extensivity makes
this fraction essentially independent of mass number.  However,
including triples excitations increases the cost to $A^3u^5$, which is
significant because $A={\cal O}(10)$ to ${\cal O}(100)$ and $u\gg A$.

To avoid this problem, several triples approximations have been
introduced over the years, see,
e.g. Refs.~\cite{lee1984,urban1985,kucharski1986,noga1987b,deegan1994,kowalski2000,piecuch2002,taube2008}. These
approaches reduce the computing (and sometimes also storage) demands
by expressing the triples amplitudes in terms of known quantities or
by including only a subset of diagrams in their computation. They all
aim at computing the energy gain from triples excitations included in
the wave function.

Here, we propose a different path that focuses on shifting the effects
of triples excitations from the wavefunction to the Hamiltonian. This
approach seems particularly attractive in nuclear physics where one
deals with Hamiltonians containing two- and three-nucleon
interactions. These are resolution-scale
dependent~\cite{bogner2003,nogga2004,bogner2007,jurgenson2009,bogner2010,hebeler2011},
i.e. they depend on an arbitrarily chosen dividing scale (i.e. the
high-momentum cutoff $\Lambda$) that separates resolved long-range
physics from unresolved (and unknown) short-range stuff. However,
low-energy observables are resolution-scale independent and the change
of the resolution (or renormalization) scale can be viewed as a
similarity transformation~\cite{bogner2007}. Such transformations
shift physics from the Hamiltonian to the wavefunction (and vice
versa). We mention several examples. \textcite{lepage1997} showed how
the removal (``integrating out'') of short-range physics involving
momenta larger than a given cutoff $\Lambda$ can be compensated by
renormalization using a short-range interaction of physical range
$1/\Lambda$ or smaller. This is beautifully demonstrated in similarity
renormalization group transformations of light
nuclei~\cite{jurgenson2009,jurgenson2011,neff2015}, in the
resolution-scale dependent interpretations of electron-nucleon
scattering
experiments~\cite{anderson2010,bogner2012,weiss2015,chen2017,more2017,weiss2018,tropiano2021},
and in the computation of the Gamow-Teller decay of $^{100}$Sn with
interactions and two-body currents from chiral effective field
theory~\cite{gysbers2019}.

This motivates us to think about short-range correlations in the
coupled-cluster state $|\psi\rangle$.  The CCSD approximation
introduces two-body correlations, and this in particular includes
short-range two-body correlations. Thus, the CCSD wavefunction is
accurate when two particles come close to each other, but still
further apart than the distance $1/\Lambda$. (Here, we assume that the
single-particle basis is sufficiently large and exhibits an
ultraviolet cutoff $\Lambda_{\rm UV}\gtrsim
\Lambda$~\cite{konig2014}.)  However, the CCSD approximation becomes
inaccurate if three (or more) particles are close. The inclusion of
triples excitations would remedy this
shortcoming. \textcite{lepage1997} taught us that one deals with this
problem by adding a short-range three-body interaction with a suitably
chosen strength such that it renormalizes the Schr{\"o}dinger
equation.  In other words, the CCSD approximation removes (or
excludes) short-range physics in the three-body sector from the wave
function. This then requires the renormalization of the Hamiltonian,
which in this case introduces a short-range three-body potential.

To see this, we consider the coupled-cluster energy 
\begin{align}
    E &=\langle\phi|e^{-T_1-T_2-T_3}He^{+T_1+T_2+T_3}|\phi\rangle\nonumber\\
    &=\langle\phi|e^{-T_1-T_2}\left(e^{-T_3}H e^{T_3}\right)e^{+T_1+T_2}|\phi\rangle
\end{align}
in the singles, doubles, and triples approximation.  Here we shifted
the $T_3$ correlations from the wavefunction to the Hamiltonian. Let
us now assume that the main effects of triples $T_3$ consist of
short-ranged three-body correlations. Then, following
Ref.~\cite{lepage1997},
\begin{align}
\label{master}
    e^{-T_3}H e^{T_3} \approx H +V_3 \ .
\end{align}
Here, $V_3$ denotes a three-body contact. The relation~(\ref{master})
is not an operator identity (the right-hand side is Hermitian, while
the left-hand side is not) but rather a low-energy (or long
wavelength) approximation. Systematic corrections consist of
derivatives acting on the contact, see Ref.~\cite{lepage1997}. In what
follows, we will limit ourselves to the leading contact.

We note here that the extension of Lepage's argument from two- to
three-body systems becomes also clear when using hyperspherical
coordinates. Then, a three-body collision is clearly short ranged as
the hyperradius becomes small, and this physics -- when integrated out
by lacking wavefunction correlations -- must be included by
renormalizing a hyperspherical contact. This corresponds then to a
three-body contact in single-particle coordinates.

We see now why this renormalization is particularly attractive in
nuclear physics. Here, a three-body contact already appears at leading
order in pion-less effective field theory~\cite{bedaque1999} and at a
next-to-leading (next-to-next-to-leading) order in chiral effective
field with (without) delta
isobars~\cite{vankolck1994,epelbaum2002}. Thus, restricting the
computational solution of the nuclear many-body problem to CCSD simply
requires one to renormalize the strength of that contact.

{\it Renormalization of the three-body contact.---}
We employ the nuclear Hamiltonian
\be
H = T_{\rm in} + V_{NN}  + V_{NNN} \ .
\ee
Here, $T_{\rm in}$ denotes the intrinsic kinetic energy (i.e. the
total kinetic energy minus that of the center of mass), $V_{NN}$ the
nucleon-nucleon interaction, and $V_{NNN}$ the three-nucleon
potential. The coupled-cluster computations start from the
Hartree-Fock basis, and the Hamiltonian is normal-ordered with respect
to the Hartree-Fock reference state. Following the normal-ordered
two-body approximation~\cite{hagen2007b,roth2012}, we neglect the
residual three-body interaction.

We employ two interactions, namely 1.8/2.0(EM) from
Ref.~\cite{hebeler2011} (labelled as interaction~A) and
$\Delta$NNLO$_{\rm GO}$(394) from Ref.~\cite{jiang2020} (labelled as
B).  We renormalize their three-body contact $c_E$ in $^{16}$O,
requiring that CCSD computations of the ground-state energies with the
renormalized interactions agree (to four significant digits) with
triples results using the original interactions. For the triples
computations we use $\Lambda$-CCSD(T)~\cite{taube2008} for
interaction~A and CCSDT-1 for interaction~B (taken from
Ref.~\cite{jiang2020}).  Table~\ref{tab:interactions} shows the
renormalized values of $c_E$ and compares them with the original ones.
In our computations we use a model space consisting of 15 harmonic
oscillator shells with a frequency $\hbar\omega=16$~MeV.

\begin{table}[htb]
\caption{
\label{tab:interactions}
Employed interactions are 1.8/2.0(EM) from Ref.~\cite{hebeler2011}
(labelled as A) and $\Delta$NNLO$_{\rm GO}$(394) from
Ref.~\cite{jiang2020} (labelled as B). Their renormalized versions
only differ by the modified three-body contact $c_E$ from the
originals.  }
\begin{ruledtabular}
\begin{tabular}{l|l|d|}
Interaction & Name & c_E \\ \hline
A & \multirow{2}{*}{1.8/2.0(EM)} & -0.12~\mbox{\cite{hebeler2011}}\\
A renorm. &  & -0.0665\\\hline
B & \multirow{2}{*}{$\Delta$NNLO$_{\rm GO}$(394)} & -0.002~\mbox{\cite{jiang2020}} \\
B renorm. &  & 0.11\\
\end{tabular}
\end{ruledtabular}
\end{table}

We turn to computations of other nuclei, performing CCSD computations
with the properly renormalized interactions. Results are shown in
Table~\ref{tab:nuclei}. The CCSD results based on the renormalized
interactions are very close to the triples results, with the largest
deviation (in $^{40}$Ca for interaction B) being less than 2\%. This
demonstrates that triples indeed account mainly for short-ranged
three-body correlations, and that the proposed renormalization is
effective.

\begin{table}[hbt]
\caption{
\label{tab:nuclei}
Binding energies (in MeV) for selected nuclei computed with CCSD using
the renormalized interactions and compared to triples results
[$\Lambda$-CCSD(T) for interaction~A and CCSDT-1 for interaction~B]
using the original interactions. Experimental values are shown in the
last column.}
\begin{ruledtabular}
\begin{tabular}{c|l|l|l|l|r }
& \multicolumn{4}{c|}{Interaction and method} & \\
&  A renorm. & A  & B renorm. & B &  Exp.\\
& {CCSD}     & {$\Lambda$-CCSD(T)} & {CCSD} & {CCSDT-1} & \\
\colrule
$^{16}$O   & 127.8 & 127.8 & 127.5 & 127.5 & 127.62 \\
$^{24}$O   & 166   & 165   & 169   & 169   & 168.96 \\
$^{40}$Ca  & 346   & 347   & 341   & 346   & 342.05 \\
$^{48}$Ca  & 420   & 419   & 419   & 420   & 416.00 \\
$^{78}$Ni  & 642   & 638   & 636   & 639   & 641.55 \\
$^{90}$Zr  & 798   & 795   & 777   & 782   & 783.90 \\
$^{100}$Sn & 842   & 836   & 816   & 818   & 825.30 \\
\end{tabular}
\end{ruledtabular}
\end{table}

\begin{figure}[!htbp]
    \includegraphics[width=0.99\linewidth]{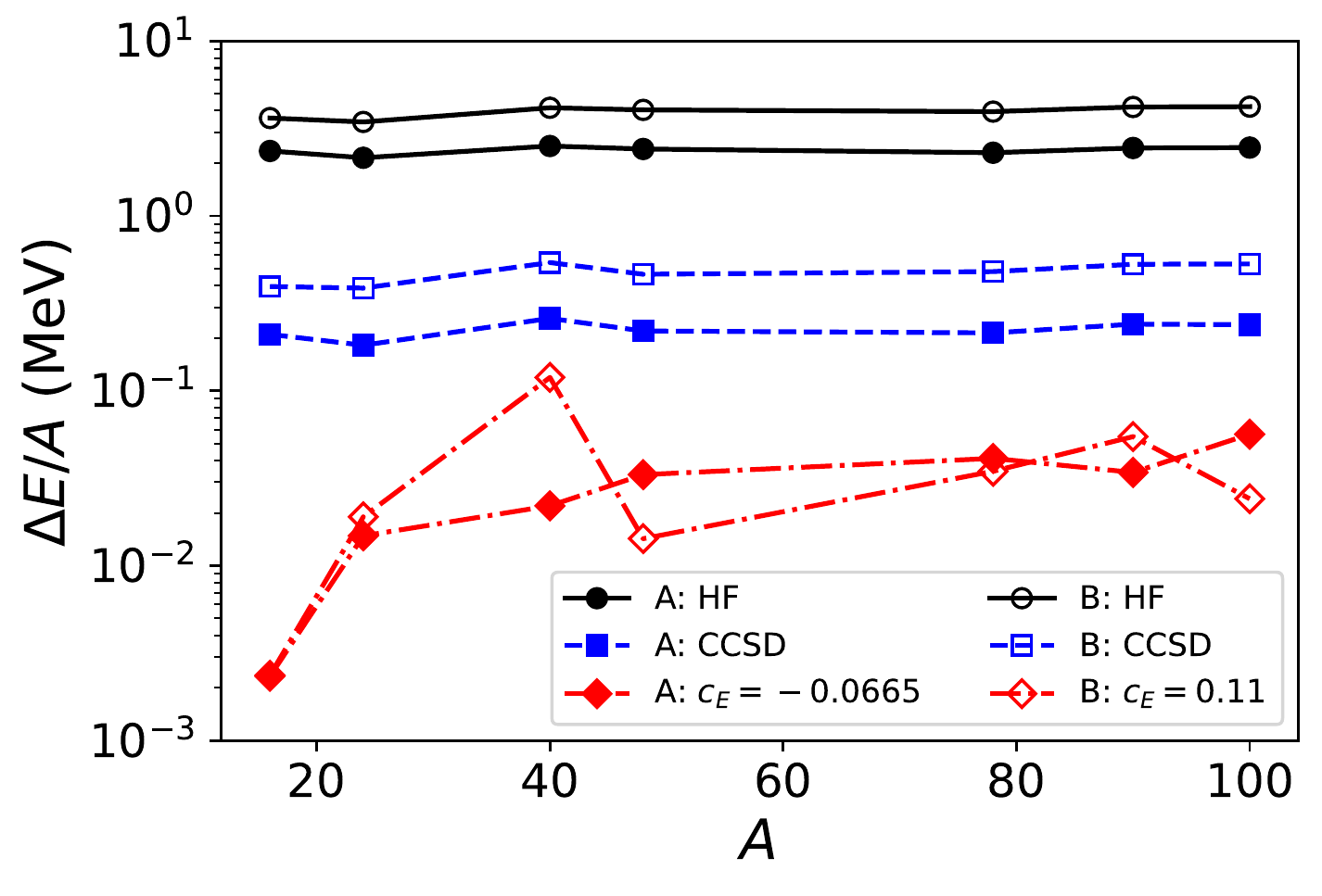}
    \caption{Absolute difference of energies per nucleon with respect
      to triples benchmarks as a function of nucleon number for
      $^{16,24}$O, $^{40,48}$Ca, $^{78}$Ni, $^{90}$Zr, and $^{100}$Sn
      using Hartree Fock (HF, black circles) and CCSD (blue squares)
      for the interactions~A (full markers) and B (hollow markers) and
      for CCSD computations with the renormalized interactions (red
      diamonds) where the three-body contact has $c_E$ as labelled.}
    \label{fig:lepage}
\end{figure}

How systematic is the improvement coming from renormalization? To
address this question, we take the triples values from
Table~\ref{tab:nuclei} as benchmarks and compute the absolute
differences (with respect to the benchmark) of the energy per particle
for Hartree Fock and for CCSD using the original interactions A and
B. We also compute the absolute differences of the CCSD energy per
particle using the renormalized interactions. The results are shown in
Fig.~\ref{fig:lepage} for interaction~A (B) as full (hollow) markers,
using black circles and blue squares for Hartree Fock and CCSD,
respectively, with the original interactions, and red diamonds for
CCSD with the renormalized interactions.  For the original
interactions, CCSD gives an order-of-magnitude improvement in accuracy
over Hartree Fock. (We also see that interaction~A is softer than
interaction B because it is closer to the triples benchmark for
Hartree Fock and CCSD.) The CCSD computations with the renormalized
interactions improve the accuracy by another order of magnitude. This
shows that the renormalization indeed yields a systematic
improvement. As already seen in Table~\ref{tab:nuclei}, the nucleus
$^{40}$Ca is a bit an outlier for interaction~B; however, the
improvement in accuracy is still about a factor of four also here.

Two comments are in order. First, changing the renormalized $c_E$
value about 5-10\% does not reduce the systematic improvement. Thus,
$c_E$ is not finely tuned (and could probably be also renormalized in
a nucleus different from $^{16}$O). Second, performing the
renormalization in $^4$He does not yield accurate results for heavier
nuclei. We attribute this to the fact that triples corrections in
$^4$He are much smaller than the usual 10\% of the correlation energy
obtained for heavier nuclei.

\begin{figure}[!htbp]
    \includegraphics[width=0.99\linewidth]{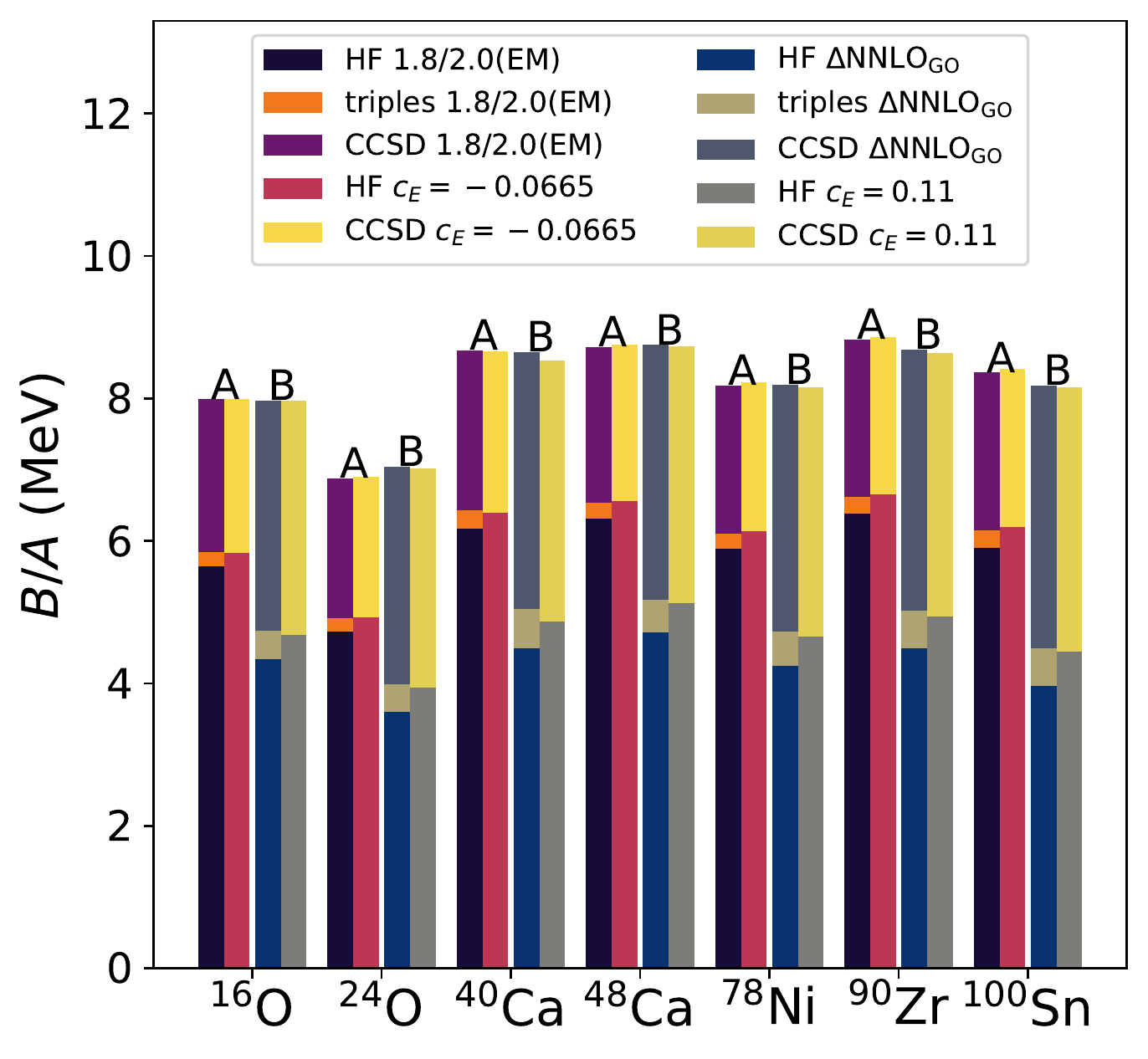}
    \caption{Energy contributions from Hartree Fock (HF), triples, and
      CCSD to the binding energy per nucleon, $B/A$, of various
      nuclei, computed with the original interactions A and B (three
      stacked left bars in pairs of columns) and compared with
      Hartree-Fock and CCSD energies from the renormalized
      interactions (two stacked right bars in pairs of columns). For
      each nucleus the left and right pairs of columns show the
      results for the interaction A and B, respectively.}
    \label{fig:Bars}
\end{figure}

How are the triples contributions accounted for in the renormalized
interaction?  We found that essentially the whole triples
contributions to the binding energies using the original interactions
become part of the Hartree-Fock energies when using the renormalized
interactions, i.e. the energy contributions from CCSD using the
original or the renormalized interactions are virtually the same.
This is shown in Figs.~\ref{fig:Bars} for both interactions.

{\it Power counting.---} The approach via renormalization makes it
clear how one would further improve these results, i.e. bring the CCSD
calculations with renormalized interactions closer to the triples
benchmarks~\cite{lepage1997}: The subleading corrections consist of
three-body contact terms with two derivatives. \textcite{girlanda2011}
showed that there are 13 such terms with different spin-isospin
structures, and this would require one to adjust as many low-energy
coefficients to data. While one could, for example, accomplish this in
mass-table computations, such an approach is beyond the scope of this
work. The important point here is that renormalization offers us a way
to systematically improve the results. The key question then concerns
the power counting, i.e. by how much would one expect the subleading
corrections to get closer to the triples benchmark?

The proposed renormalization scheme must break down when triples
correlations are not dominantly short ranged.  A derivative on the
three-body contact yields a momentum ${\rm min}(k_3,k_{\rm typ})$
where $k_3$ is a small three-body momentum (because we lack
short-ranged three-body correlations) and $k_{\rm typ}$ is the typical
momentum which could be of the scale of the Fermi momentum. We have
$k_3\ll k_{\rm typ}$ when three-body correlations are short
ranged. The derivative's contribution fails to be small (compared to
the leading three-body contact) if $k_3\approx k_{\rm typ}$, i.e. for
low-density nucleons without short-range three-body correlations. We
therefore propose that the power counting is in the ratio $k_3/k_{\rm
  typ}$. This ratio must be small for nuclei, because the leading
contact recovers so much of the triples benchmarks.

\begin{figure}[!htbp]
    \includegraphics[width=0.99\linewidth]{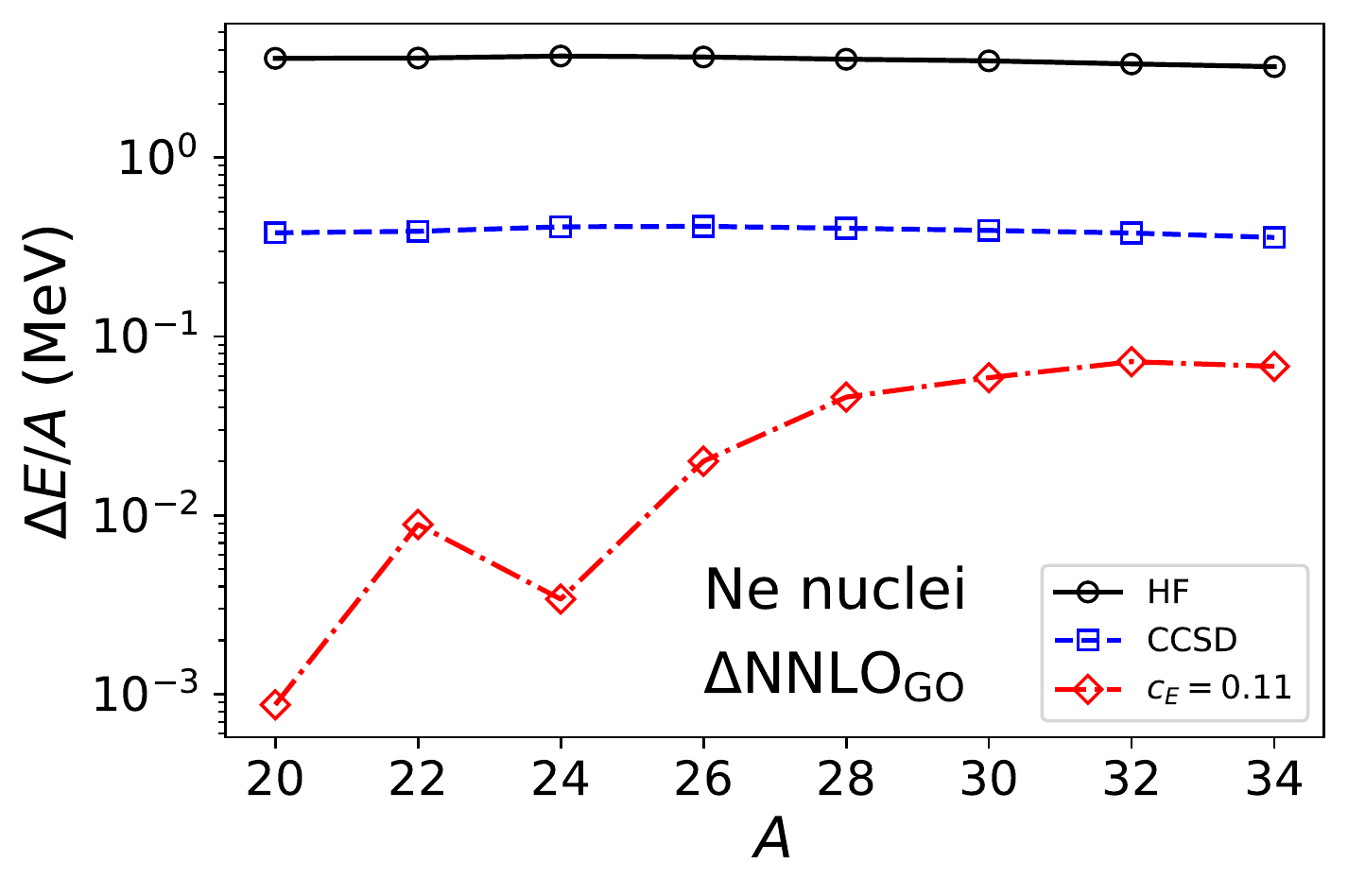}
    \caption{Absolute difference of energies per nucleon with respect
      to CCSDT-1 as a function of nucleon number for $^{20-34}$Ne
      nuclei using Hartree Fock (HF, black circles) and CCSD (blue
      squares) for the interaction~B and for CCSD computations with
      the renormalized interaction where the three-body contact has
      been renormalized with $c_E=0.11$ (red diamonds).}
    \label{fig:neon}
\end{figure}

The arguments proposed below would entail that the renormalization is
less effective in low-density matter. This makes it interesting to
study dripline nuclei. Using the renormalized interaction B, we also
computed neutron-rich neon isotopes and compared with the triples
results~\cite{novario2020} of the original interaction. The
calculations are based on an axially-symmetric deformed Hartree-Fock
state, and they lack angular momentum projection. The results, shown
in Fig.~\ref{fig:neon}, demonstrate that the renormalization
significantly and systematically improves the ground-state
energies. For the most neutron-rich isotopes, though, the accuracy is
``only'' improved by a factor of about four. The trend of reduced
gains from renormalization as the the dripline is approached is
consistent with the arguments made for the power counting. Thus, our
calculations of neon nuclei show that the renormalization proposed in
this work is also useful for open-shell nuclei.

We also computed the charge radius of $^{20}$Ne and found that the
renormalized interaction yields about 1.8\% less than the original
one~\cite{novario2020}. This is consistent with what is found for the
RG evolution of long-ranged operators~\cite{schuster2014}.

\begin{figure}[!htbp]
    \includegraphics[width=0.99\linewidth]{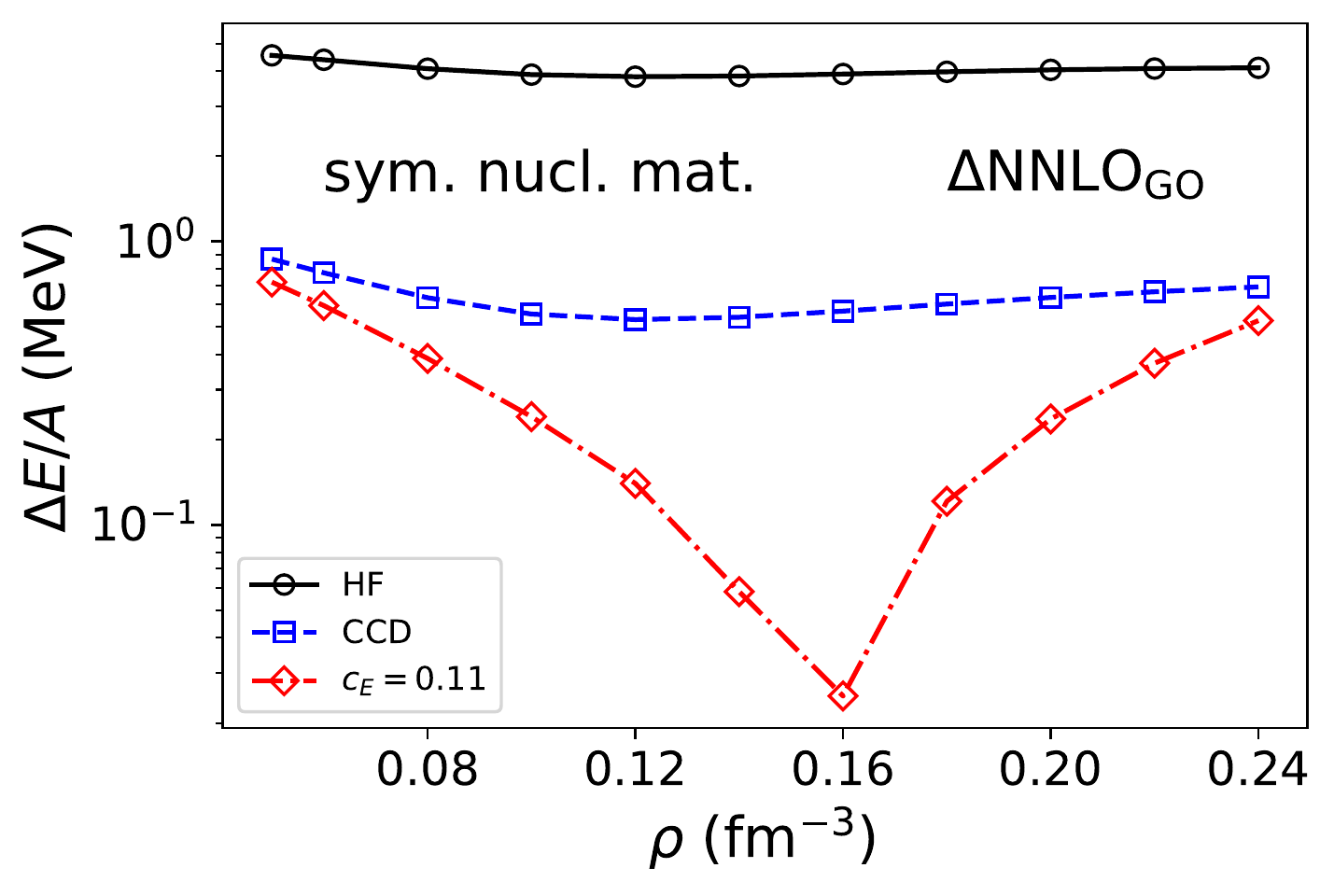}
    \caption{Absolute difference of energies per nucleon with respect
      to CCD(T) as a function of density using Hartree Fock (HF, black
      circles) and CCD (blue squares) for the interaction~B and for
      CCD computations with the renormalized interaction where the
      three-body contact has been renormalized with $c_E=0.11$ (red
      diamonds).}
    \label{fig:snm}
\end{figure}

We finally turn to symmetric nuclear matter and perform the
computations following Refs.~\cite{hagen2013b,jiang2020}, taking
CCD(T) as the benchmark. The calculations use $A=132$ nucleons on a
momentum-space lattice (with $n_{\rm max}=4$) corresponding to
periodic boundary conditions in position space. We checked that
CCSDT-1 benchmarks are close to the less expensive CCD(T) for
$A=28$. Figure~\ref{fig:snm} shows the absolute difference to the
triples benchmark of the energy per nucleon as a function of the
density $\rho$ for Hartree Fock and CCD with the interaction B, and
for CCD with the renormalized interaction B. We see that CCD with the
interaction renormalized in $^{16}$O is very accurate around
saturation density. We note that the energy difference changes sign
there. Inspection also shows that the Hartree-Fock energy for the
renormalized interaction differs from that of the original one by a
contribution proportional to $\rho^2$. This explains the trend seen
for neutron-rich neon nuclei.  Figure~\ref{fig:snm} also shows that
the renormalization breaks down at low densities ($\rho\approx
0.06$~fm$^{-3}$) and high densities ($\rho\approx 0.24$~fm$^{-3}$).

{\it Discussion and summary.---} We have seen that the extensive
energy contributions from nuclear three-nucleon correlations can be
captured in CCSD via a renormalization of the three-body contact. Our
results are based on (and consistent with) the assumption that
three-nucleon correlations are dominantly short ranged. This suggests
that arguments about the universality of short-range two-body
correlations~\cite{konig2014,weiss2015,weiss2018,tropiano2021} extend
to three-nucleon correlations.

While our discussions focused on the coupled-cluster theory, this
method is closely related to the in-medium similarity renormalization
group (IMSRG)~\cite{tsukiyama2011,hergert2016,stroberg2017}, Green's
function approaches~\cite{dickhoff2004} and Gorkov
methods~\cite{soma2013}. In the IMSRG, for instance, capturing
three-body correlations comes at a very high
cost~\cite{heinz2021}. Including three-body correlations in the trial
wave functions of variational Monte Carlo is also a challenging
task~\cite{wiringa2000,adams2021}. This suggests that the
renormalization proposed in this paper could also be useful for these
methods.

The insights presented in this paper also explain why triples
correlations play a smaller role in neutron matter~\cite{hagen2013b}
than in nuclear matter: the Pauli principle prevents short-ranged
three-neutron correlations, and the leading renormalization comes from
terms where two derivatives act on a three-body contact.

One might also consider to take the renormalization proposed in this
paper to its extreme: Hartree-Fock computations even exclude two-body
correlations. This suggests that one could also employ Hamiltonians
from effective field theories using properly renormalized two- and
three-body contacts (and derivatives acting on them). This somewhat
resembles the expansions~\cite{raimondi2014,navarro2018} for density
functionals but would be for Hamiltonians~\cite{schmidt1989}.

The proposed renormalization scheme significantly lowers the
computational cost for nuclear binding energies and thereby puts
Hamiltonian-based masstable computations of atomic
nuclei~\cite{stroberg2021} in closer reach of various {\it ab initio}
methods.  It also links correlations in many-body systems to the
renormalization group and thereby offers new ways to think about their
role.

\begin{acknowledgments}
We thank C.~Forss{\'e}n, R.~J.~Furnstahl, H.~Hergert, T.~D.~Morris,
and D.~R.~Phillips for useful discussions.  This work was supported by
the U.S. Department of Energy, Office of Science, Office of Nuclear
Physics, under Award Nos.~DE-FG02-96ER40963 and DE-SC0018223, and by
the Quantum Science Center (QSC), a National Quantum Information
Science Research Center of the U.S. Department of Energy (DOE).
Computer time was provided by the Innovative and Novel Computational
Impact on Theory and Experiment (INCITE) programme. This research used
resources of the Oak Ridge Leadership Computing Facility located at
Oak Ridge National Laboratory, which is supported by the Office of
Science of the Department of Energy under contract
No. DE-AC05-00OR22725.
\end{acknowledgments}

\end{document}